\documentclass[10pt,aps,prl,groupedaddress,twocolumn,amsmath,amssymb]{revtex4-1}

\usepackage{nicefrac,bm,mathtools}                      
\usepackage{graphicx}                                   
\usepackage[pdftex]{hyperref}                           
\usepackage{siunitx}
\usepackage[version=4]{mhchem}
\usepackage{xr}
\usepackage{verbatim}
\usepackage{xcolor}

\newcommand{\figref}[1]{Fig.~\ref{fig:#1}}

\newcommand{\subfig}[1]{\textbf{(#1)}}

\newcommand{\dd}[1]{\mathrm{d}#1}
\newcommand{\ddd}[2]{\ensuremath{\operatorname{d}^{#2}\!{#1}}}
\newcommand{\pdb}[2]{\frac{\partial #1}{\partial #2}}
\newcommand{\punc}[1]{\,#1}

\newcommand{\aR}{\alpha_{\mathrm{R}}}
\newcommand{\kR}{k_{\mathrm{R}}}

\newcommand{\eF}{\epsilon_{\mathrm{F}}}
\newcommand{\eFpm}{\epsilon^\pm_\mathrm{F}}

\newcommand{\ek}{\epsilon_k}
\newcommand{\evk}{\epsilon_{\mathbf{k}}}

\newcommand{\vk}{\mathbf{k}}
\newcommand{\rso}{\mathbf{r}_{\mathrm{SO}}}
\newcommand{\muB}{\mu_{\mathrm{B}}}

\newcommand{\Tc}{T_{\mathrm{c}}}
\newcommand{\Bc}{B_{\mathrm{c}}}
\newcommand{\sxx}{\sigma_{\mathrm{xx}}}

\newcommand{\Rxx}{R_{\mathrm{xx}}}
\newcommand{\Rxy}{R_{\mathrm{xy}}}
\newcommand{\Rmin}{R_\mathrm{min}}
\newcommand{\RN}{R_{\mathrm{N}}}
\newcommand{\dBbydt}{\nicefrac{\dd{B}}{\dd{t}}}

\begin{document}

\title{Long-lived non-equilibrium superconductivity in a non-centrosymmetric Rashba semiconductor}

\author{V. Narayan$^{1,*}$, P. C. Verpoort$^1$, J. R. A. Dann$^1$, D. Backes$^{1}$, C. J. B. Ford$^1$, M. Lanius$^3$, A. R. Jalil$^3$, P. Sch{\"u}ffelgen$^3$, G. Mussler$^3$, G. J. Conduit$^1$, D. Gr{\"u}tzmacher$^3$}
\affiliation{
  $^1$Department of Physics, University of Cambridge, J.J. Thomson Avenue, Cambridge CB3 0HE, UK. \\
  $^2$Department of Materials Science \& Metallurgy, University of Cambridge, J.J. Thomson Avenue, Cambridge CB3 0FS, UK. \\
  $^3$Peter Gr\"unberg Institute (PGI-9), Forschungszentrum J\"ulich \& J\"ulich-Aachen Research Alliance (JARA-FIT), 52425 J\"ulich, Germany. \\
  $^\dagger$Current address: Department of Physics, Loughborough University, Loughborough LE11 3TU, UK. \\
  $^*$Correspondence to: V. Narayan $\langle$\href{mailto:vn237@cam.ac.uk}{\nolinkurl{vn237@cam.ac.uk}}$\rangle$
}

\date{\today}

\begin{abstract}
We report non-equilibrium magnetodynamics in the Rashba-superconductor GeTe, which lacks inversion symmetry in the bulk. We find that at low temperature the system exhibits a non-equilibrium state, which decays on time scales that exceed conventional electronic scattering times by many orders of magnitude. This reveals a non-equilibrium magnetoresponse that is asymmetric under magnetic field reversal and, strikingly, induces a non-equilibrium superconducting state distinct from the equilibrium one. We develop a model of a Rashba system where non-equilibrium configurations relax on a finite timescale which captures the qualitative features of the data. We also obtain evidence for the slow dynamics in another non-superconducting Rashba system. Our work provides novel insights into the dynamics of non-centrosymmetric superconductors and Rashba systems in general.
\end{abstract}


\maketitle

Rashba systems are a class of spin-orbit coupled materials in which spatial inversion symmetry is absent and whose band structure, therefore, lacks spin degeneracy. The dispersion of Rashba systems features two concentric Fermi surfaces with opposing helical spin structures that are separated in momentum space by twice the Rashba wavevector $\kR$. Systems displaying a large Rashba effect are desirable for all-electrical spin-based logic schemes, and Rashba superconductors are expected to harbour topological superconducting phases~\cite{Sato09}, much sought-after towards fault-tolerant quantum computation. While the breaking of spatial inversion is most readily achieved in low-dimensional systems, recently three-dimensional materials such as bismuth tellurohalides~\cite{Bahramy17} and GeTe~\cite{Sante12} have been shown to have a giant bulk Rashba effect.

It is known that the presence or absence of specific symmetries in a system has a telling effect on the allowed dynamical processes~\cite{Narayan07}. In the specific case of Rashba systems, transitions between the two Rashba bands are constrained by the finite momentum split $\kR$ and the helical spin-structure. As we will show in this Letter, this has important consequences for equilibration. The spin-structure of the Rashba bands also has important consequences for superconducting systems and, in particular, the nature of Cooper pairs~\cite{Bauer12,Gorkov01}. Thus, Rashba superconductors can harbour unconventional superconducting phases including Fulde-Ferrell-Larkin-Ovchinnikov-type~\cite{Fulde64,Larkin65} phases in which the Cooper pair has a finite momentum, and/or topological superconductor phases~\cite{Sato09, Tafti_PRB2013, Nakajima_SciAdv2015, Sato17, Xiao_etal_PRB2018}.

\begin{figure*}[t]
  \includegraphics[width=14cm]{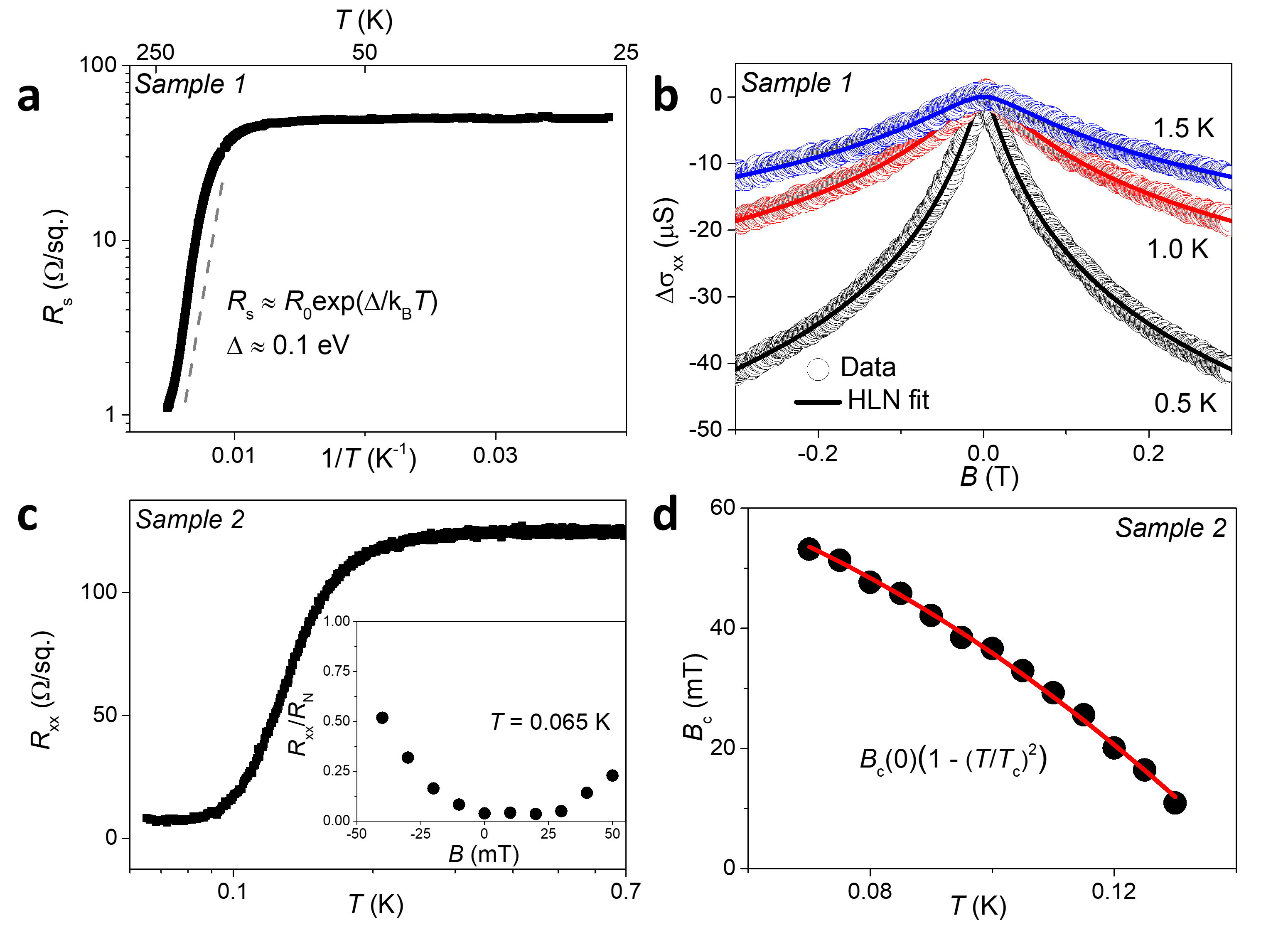}
  \caption{GeTe: a semiconducting, spin-orbit-coupled superconductor. \subfig{a} The ultra-thin GeTe films shows semiconducting characteristics at high $T$ with a band gap of $\approx \SI{0.1}{\electronvolt}$. Below \SI{100}{\kelvin}, the $T$ dependence weakens significantly, being indicative of 2D metallic states. \subfig{b} The spin-orbit coupling in GeTe manifests as WAL, i.e., positive quantum corrections to the electrical conductivity $\Delta\sxx = \sxx(B)-\sxx(0)$. These characteristics are well described by the Hikami-Larkin-Nagaoka (HLN)~\cite{Hikami80} formula valid for 2D systems (see also Fig.~S2~\cite{SOM}). \subfig{c} There is a broad superconducting transition between \SI{0.2}{\kelvin} and \SI{0.1}{\kelvin} below which we note that $\Rxx$ does not go completely to zero (see Fig.~S3~\cite{SOM} for possible explanation). Inset: superconductivity is suppressed when the sample is cooled in the presence of a constant magnetic field perpendicular to the plane of the film. The data is consistent with a field offset of $\approx\SI{15}{\milli\tesla}$, which can arise due to trapped flux in the external superconducting magnet. \subfig{d} The dependence of the critical field $\Bc$ on $T$ reveals $\Bc(\SI{0}{\kelvin}) = \SI{70}{\milli\tesla}$ and $\Tc = \SI{140}{\milli\kelvin}$. Here the superconducting transition is defined as $\Rxx < \RN/2$ (where $\RN$ is the normal-state resistance).}
  \label{fig:fig02}
\end{figure*}


We report here low temperature ($T$), magnetic field ($B$)-induced dynamics in molecular-beam-epitaxy (MBE)-grown (Fig.~S1~\cite{SOM}) ultra-thin films (\SI{18}{nm}-thick) of GeTe. GeTe is a narrow band-gap semiconductor with giant bulk and surface Rashba couplings~\cite{Sante12,Liebmann16}, and is inherently superconducting~\cite{Hein64,Narayan16}. We present data from two Hall bar samples patterned from the same wafer (for details see \cite{SOM}), which go superconducting below \SI{0.2}{\kelvin}. Strikingly, we find that a second non-equilibrium superconducting state with a higher critical temperature ($\Tc$) and critical field ($\Bc$) is accessed when the system is subjected to a continuously-ramped magnetic field $B(t)$. This non-equilibrium state is extremely long-lived, relaxing on macroscopic timescales of several minutes. By ruling out other well-known sources of slow dynamics, we demonstrate that the mechanism underlying the observed dynamics is novel. We have shown that such long-lived non-equilbrium behaviour can generically be expected in clean Rashba systems~\cite{our_paper_theory}: in materials with strong Rashba coupling, where $\kR$ is much larger than the thermal phonon momentum scale, there is a suppression of all relaxation processes involving real phonon modes. Furthermore, the spin texture at the Fermi surface serves to significantly reduce the scattering events due to inter-carrier interactions, ultimately resulting in non-equilibrium states with finite lifetimes. Based on this we formulate a model of a Rashba superconductor with suppressed inter-band transitions, within which the non-equilibrium superconducting behaviour arises from the enhancement of the density of states. Our model suggests similar dynamics in normal Rashba materials, evidence of which we observe in a topological insulator heterostructure with dominantly bulk-type transport.

\begin{figure*}
  \includegraphics[width=\textwidth]{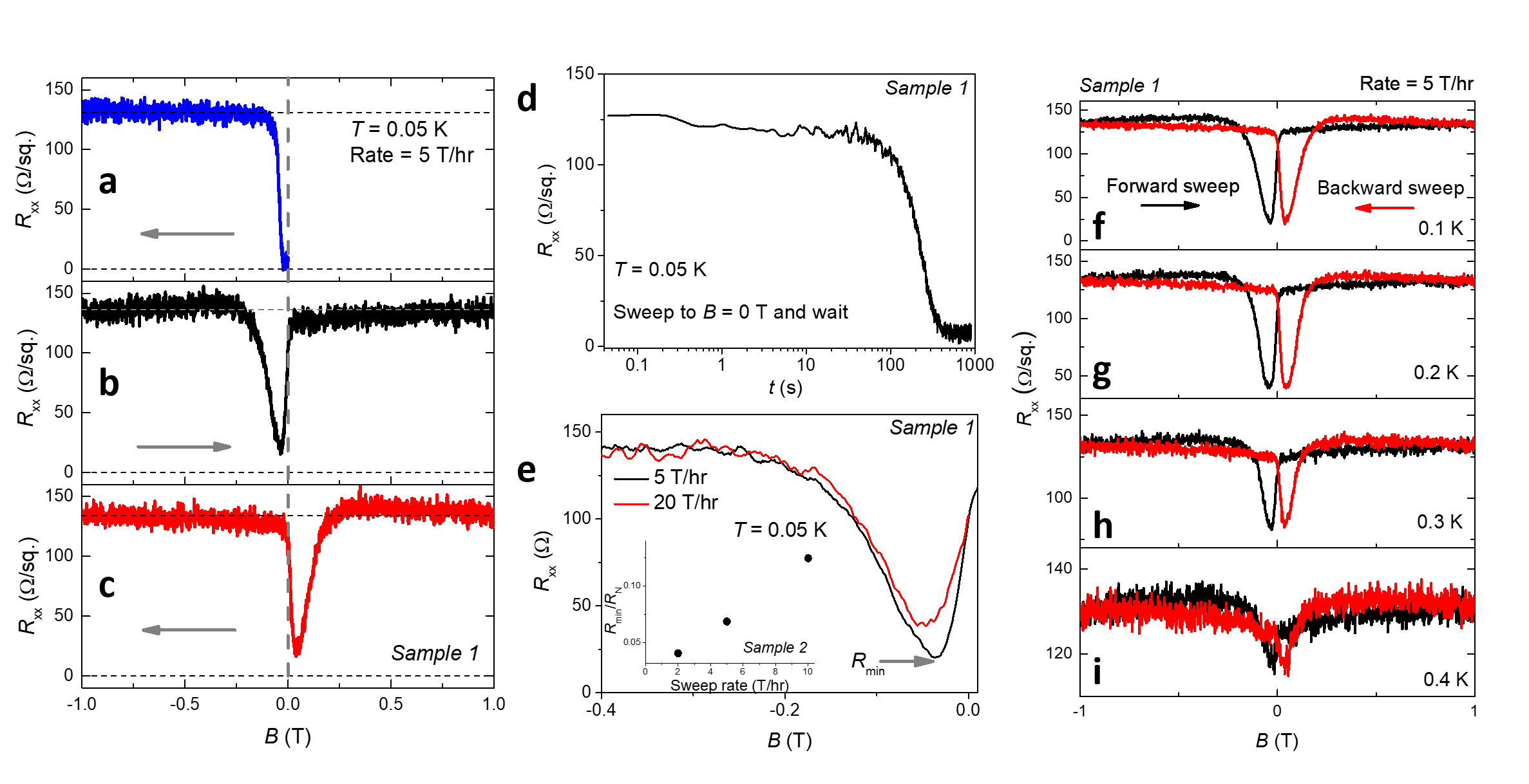}
  \caption{Long-lived non-equilibrium magnetodynamics. \subfig{a} Starting in the superconducting state, as $B$ is ramped towards \SI{-1}{\tesla} at $\dBbydt = \SI{5}{\tesla\per\hour}$ (\SI{1.4}{\milli\tesla\per\second}) there is a sharp transition to the normal state at \SI{-15}{\milli\tesla}. However, the `forward $B$ sweep` from \SI{-1}{\tesla} to \SI{+1}{\tesla} shown in \subfig{b} is distinctly dissimilar to the previous trace, showing an initial increase in $\Rxx$ followed by an almost complete transition to superconductivity beginning at $B = \SI{-250}{\milli\tesla}$, well above the previously estimated values of $\Bc$. The sweep is not symmetric about $\Rmin$ or about $B = \SI{0}{\tesla}$. In contrast, at $B = \SI{0}{\tesla}$, the value $\Rxx$ is now finite and close to $\RN$. \subfig{c} The shape of $\Rxx$ as $B$ is swept in the opposite direction from \SI{+1}{\tesla} to \SI{-1}{\tesla} is precisely the mirror image of the forward sweep when reflected about $B = \SI{0}{\tesla}$. The grey arrows indicate the direction of the $B$ sweep. \subfig{d} When the sweep is stopped at $B = \SI{0}{\tesla}$ the non-equilibrium state persists for $\approx\SI{100}{\second}$ before relaxing to the equilibrium zero-resistance state over a further $\SI{300}{\second}$. \subfig{e} Over the range of sweep rates explored, $\Rxx$ is seen to show a lower minimum $\Rmin$ for slower sweep rates. Inset: $\Rmin/\RN$ as a function of sweep rate. \subfig{f} -- \subfig{i} The apparent non-equilibrium superconducting state is visible even above $\Tc$.}
  \label{fig:fig03}
\end{figure*}

\figref{fig02} characterises the equilibrium electrical properties of GeTe in the normal and superconducting states. Between room $T$ and $\approx\SI{100}{\kelvin}$ GeTe shows activated behaviour indicative of a band gap of \SI{0.1}{\electronvolt}. Below \SI{100}{\kelvin} the transport becomes $T$ independent, suggesting the role of two-dimensional (2D) modes. Correspondingly, the electrical characteristics are plotted as `sheet resistances' $R_{\mathrm{s}} = R \times W \, / \, L$, where $R$ is the electrical resistance measured using a constant-current, four-terminal setup, and $W$ and $L$ are the width and length of the Hall bar. The 2D nature of transport is corroborated by the low-$T$ magnetotransport, where positive quantum corrections to the electrical conductivity $\sxx$ or weak anti-localisation (WAL) characteristics are seen to be 2D-like (\figref{fig02}b, Fig.~S2~\cite{SOM}). Here $\sxx \equiv (L/W) \, \Rxx/(\Rxx^2 + \Rxy^2)$, where $\Rxx$ and $\Rxy$ are the longitudinal and Hall components of resistance, respectively. The observation of 2D modes is consistent with recent spectroscopic measurements~\cite{Liebmann16}, although our results are not affected by the dimensionality of transport. \figref{fig02}c shows the onset of the superconductivity at $T = \SI{0.2}{\kelvin}$ and its suppression under the influence of a perpendicular magnetic field ($B$).

Evidence of a second \textit{non-equilibrium} superconducting state is shown in \figref{fig03}. Under the influence of a slowly-ramped $B$ field perpendicular to the plane of the sample, it is found that (\figref{fig03}a-c): (1) superconductivity is no longer observed at $B = \SI{0}{\tesla}$, but instead appears at a finite $B$; (2) the magnetoresistance is asymmetric about the new superconducting state, depending explicitly on the sign of $\dBbydt$, and (3) the non-equilibrium state is highly persistent, relaxing on the timescale of minutes (\figref{fig03}d and Fig.~S4~\cite{SOM}). The occurrence of this state relies on a finite $\dBbydt$ without which one obtains the `equilibrium' magnetoresistance (\figref{fig02}c inset). Strikingly, however, \figref{fig02}e shows that sweeping at a slower rate serves to enhance the finite-$B$ minimum ($\Rmin$), implicating an optimum sweep rate at which the non-equilibrium superconducting state manifests most clearly. The enhancement of $\Bc$ evidenced in \figref{fig03}a-c in conjunction with the behaviour in \figref{fig03}d strongly suggests that the dynamical superconducting state is distinct from the initial equilibrium superconducting state, and results from a long-lived transient configuration. This is supported by \figref{fig03}f-i (and Fig.~S6~\cite{SOM}), where we find that its existence is not contingent on the equilibrium superconducting state, occurring even above $\Tc$ and remaining perceptible up to $T = \SI{0.4}{\kelvin}$. The absence of complete loss of resistance could be interpreted as a competition of the timescales for the transition into the superconducting phase, and the decay of the non-equilibrium state (see Fig.~S4~\cite{SOM}).

There are various mechanisms that are known to result in slow relaxation in solid-state systems. In the Supplement~\cite{SOM} we discuss and rule out contributions due to magnetocaoric effects, superconducting vortices, nuclear spins, trapped flux, and inhomogeneities in the GeTe film. We also note that the recent findings of `non-reciprocal transport' in Rashba systems~\cite{Wakatsuki_etal_SciAdv2017, Hoshino_etal_PRB2018} cannot explain our experimental observations as these are equilibrium effects. In particular, we find no violation of reciprocity in the equilibrium transport (\figref{fig02}c inset).

In Ref.~\cite{our_paper_theory}, we explicitly consider the relaxation dynamics in Rashba-coupled systems and establish that in the absence of charged impurities and below a characteristic $T$, non-equilibrium configurations relax on timescales that can be many orders of magnitude greater than the conventionally-observed picosecond relaxation timescales. In our MBE-grown samples, we expect the role of charged-impurity scattering to be negligible (the spontaneously-formed Ge vacancies are uncharged~\cite{Edwards05,Edwards06}, and do not contribute to Coulomb scattering), and we show that the characteristic $T$ scale can be as large as \SI{1}{\kelvin} for GeTe. In the following we verify whether the existence of such a timescale is a sufficient condition to induce the observed novel magnetoresponse by considering a model Rashba superconductor and introducing, by hand, a finite timescale $\tau$ for inter-band transitions. We estimate $\tau \approx \SI{100}{\second}$ for the GeTe films from Fig.~S5a~\cite{SOM}.

\begin{figure*}
  \includegraphics[width=\textwidth]{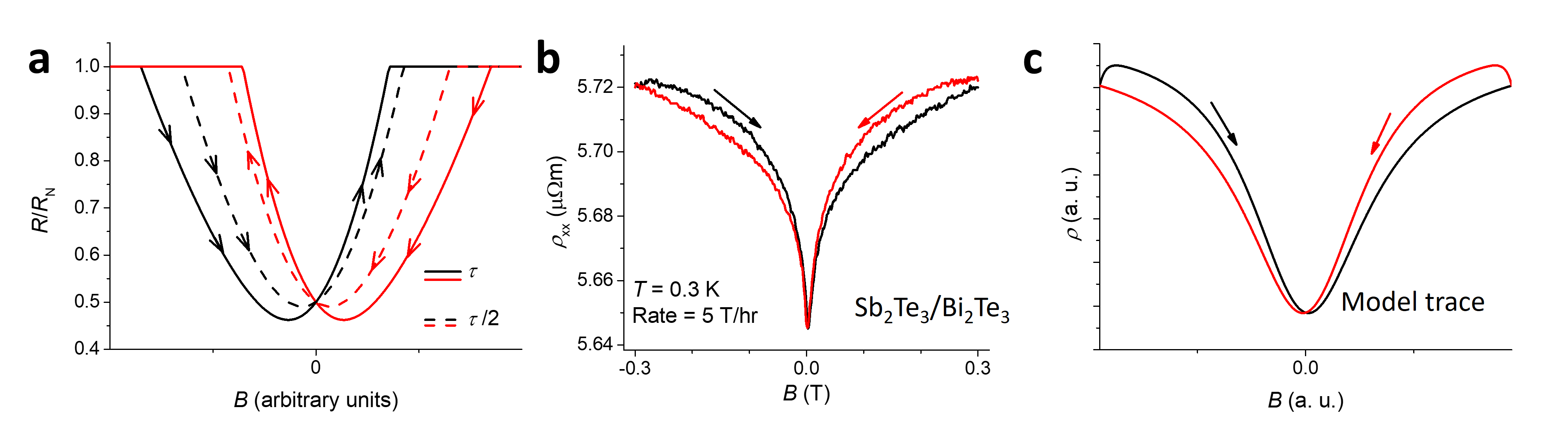}
  \caption{Non-equilibrium butterfly hysteresis. \subfig{a} The magnetoresistance of a model Rashba system subjected to a $B$ field, in which inter-band relaxation processes occur on a finite timescale $\tau$. The two traces compare the behaviour when $\tau$ is changed by a factor of $2$. Superconductivity is induced locally depending on the instantaneous carrier concentration. The model correctly captures the difference between traces with differing sign of $\dBbydt$ as well as that $R(B) \neq R(-B)$ for a given sign of $\dBbydt$. \subfig{b} The asymmetric $\rho(B)$ and mirror symmetry between the forward and reverse $B$ sweeps is seen clearly in a heterostructure of $p$ and $n$-type topological insulators. \subfig{c} The characteristic `butterfly' shape of $\rho(B)$ can be captured within a model of a Rashba material.}
  \label{fig:fig05}
\end{figure*}

The Rashba dispersion is given by $\ek^\pm = \nicefrac{\hbar^2\vk^2}{2m} \pm \sqrt{(g \muB B)^2 + (\aR\rso\times\vk)^2}$, where the $+$ ($-$) superscript refers to the inner (outer) Rashba band. Here, $\hbar$ is Planck's constant, $\vk$ is the wavevector, $m$ is the effective mass of carriers, $g$ the Lande g-factor, $\muB$ the Bohr magneton, $\aR$ the Rashba parameter, and $\rso$ is the direction of the spin-orbit coupling along which inversion symmetry is broken. This assumes the direction of the $B$ field to lie parallel to $\rso$ (however this assumption does not affect the qualitative results as in-plane fields cause a redistribution of carriers between bands similar to the out-of-plane one). To study the consequences of the time-varying magnetic field we compute the dynamical Fermi energies $\eFpm$ as a function of time $t$, whose time-dependence is governed by the differential equation:
\begin{equation}
  \frac{\dd{\eFpm}}{\dd{t}} = \pdb{\eFpm}{B} \pdb{B}{t} + \pdb{\eFpm}{n^\pm} \pdb{n^\pm}{t} \punc, \label{eq:2}
\end{equation}
where $n^\pm$ are the carrier densities of the two bands. The first term on the right describes how the Fermi surfaces change with $B$
 and the second term describes carriers relaxing so as to equilibrate $\eF^+$ and $\eF^-$. We model $\nicefrac{\partial n^\pm}{\partial t}$ using a relaxation-time approximation with time constant $\tau$.  In order to describe the superconductivity, we make the following generic assumptions: (1) superconductivity is assumed to emerge through pairing of opposite spin carriers within a band (this is the simplest prescription based on Cooper pairs with zero net momentum~\cite{Smidman_etal_RepProgPhys2017}, and while it is assumed for simplicity, the model is readily extended to include inter-band Cooper pairing); (2) the transition temperature $T_c$ is assumed to depend exponentially on the density-of-states $\nu$: $T_{c\pm} \sim \mathrm{e}^{(-1/\Gamma \nu_\pm)}$,
where the $\pm$ indicate the two Rashba bands, $\nu_\pm = \nicefrac{\dd{k}}{\dd{\epsilon_{\pm}}}$, and $\Gamma$ is the strength of the contact interaction that mediates superconductivity. $\Gamma$ is drawn from a uniform probability distribution to reflect local variations in dopant concentration, interaction strength etc.
The magnetoresistance is modelled by $R = \RN \min(T/T_{\mathrm{C-}},1)$, which arises only from the parts of the sample that are normal. The results of our model calculation are shown in \figref{fig05}a, and capture the essential features of the experimental observations in \figref{fig03}a -- c and f -- i. Notably, this simple model does not capture the $\dBbydt$-dependence of the depth of the dip (\figref{fig03}e), which is likely a higher-order effect requiring a more microscopic treatment. One plausible scenario is that the non-equilibrium superconducting state has a more complex history dependence (Fig.~S9~\cite{SOM}) or involves inter-band Cooper pairing and thus requires a threshold occupancy in both Rashba bands.

It is noteworthy that the arguments of long relaxation times presented in Ref.~\cite{our_paper_theory} are generic to Rashba systems as opposed to only Rashba superconductors. In this context, we first point out that the novel asymmetric and rate-dependent magnetodynamics are observed in GeTe even in samples that do not go superconducting~\cite{Narayan16}. \figref{fig05}b demonstrates similar, albeit less pronounced qualitative features in a vertical topological insulator (TI) $p$-$n$ junction~\cite{Eschbach15,Backes17} in which a \SI{6}{\nano\meter} layer of $\mathrm{Bi}_2\mathrm{Te}_3$ is capped with a \SI{15}{\nano\meter} layer of $\mathrm{Sb}_2\mathrm{Te}_3$. Here transport is in the plane of the film and $B$ is out of plane. TIs are well-known spin-orbit materials, and the in-built potential of the $p$-$n$ junction~\cite{Eschbach15} provides a clear mechanism for the breaking of inversion symmetry. Importantly, the specific layer configuration is known to show significant bulk transport~\cite{Backes17} rather than the surface-dominated transport in ultra-thin TIs, where this effect is not expected due to the single Dirac cone. We stress that this behaviour should not be confused with the so-called `butterfly hysteresis' observed in magnetic Dirac materials~\cite{Wolgast15,Brinkman07,Nakajima15,Daptary17,Tiwari17} since 1) the materials reported here are manifestly non-magnetic, and 2) the data reflect non-equilibrium states in the samples.

\figref{fig05}c shows the qualitative behaviour of $\rho(B)\approx\nicefrac{1}{(\sigma^+ + \sigma^-)}$ as derived from the conductivities $\sigma^\pm \propto \int \ddd{k}{3} \, (\nicefrac{\partial\evk^\pm}{\partial k})^2 \, \delta(\evk^\pm-\eFpm)$ of the bands and the dynamics of $\eFpm(t)$ arising due to Eq. \eqref{eq:2}, evaluated for a parabolic band dispersion. The figure clearly captures the salient features of the experimental data (although for a quantitative comparison, in addition to realistic band structure, one also needs to account for WAL corrections to the conductivity). The physical mechanism is the exact same as before: the two Rashba bands develop unequal Fermi levels when subjected to a $B$ sweep. Since changes in conductivity of the two bands do not cancel, i.e. $\pdb{\sigma^+}{n^+} \neq \pdb{\sigma^-}{n^-}$, there is a net change in the total conductivity $\sigma^+ + \sigma^-$. Lastly, we stress that while GeTe shows a dramatic magnetoresponse, this can be much more subtle as shown in \figref{fig05}b. We question whether similar observations may have been overlooked in the past.

In conclusion, we have reported ultra-slow relaxation and rich non-equilibrium magnetodynamics in the non-centrosymmetric Rashba superconductor GeTe. These novel dynamics reveal a second non-equilibrium superconducting state with a higher $\Tc$ and $\Bc$. Importantly, the observed slow dynamics are inconsistent with more common sources of slow dynamics such as nuclear spin relaxation and vortex creep. They are also inconsistent with magnetocaloric-driven cooling and/or Eddy current-induced heating. We develop a model that successfully captures the salient features of the experimental data, and also predicts a specific response in normal Rashba systems, qualitative evidence of which we observe in a TI-based heterostructure. We suggest, therefore, that our observations might be generic to Rashba systems and discuss the conditions under which they might be observed in experiment. Our work has provided striking new experimental insights into the behaviour of Rashba superconductors and possibly indicates a non-equilibrium behaviour unique to Rashba systems in general.

\begin{acknowledgments}
VN, JRAD, PCV, DB, and CJBF acknowledge funding from the Engineering and Physical Sciences Research Council (EPSRC), UK. GJC acknowledges funding from the Royal Society, UK. GM, ML, ARJ, PS and DG acknowledge financial support from the DFG-funded priority programme SPP1666 as well as from the Helmholtz Association via the `Virtual Institute for Topological Insulators` (VITI). VN acknowledges useful discussions with Mark Blamire, Siddharth Saxena, Niladri Banerjee, Peter Wahl, and Yoichi Ando. VN also thanks Seamus Davis for suggesting the measurements in Fig.~S8~\cite{SOM}. PCV acknowledges useful discussions with Giulio Schober. Supporting data for this publication can be found at \url{https://doi.org/10.17863/CAM.22843}.

Author contribution: VN fabricated and measured the GeTe devices, and wrote the paper with inputs from PCV, JRAD, CJBF, GM, and GJC. DB fabricated and measured the topological insulator devices. ML, ARJ, PS, GM, and DG grew and characterised the GeTe films and topological insulator heterostructures. PCV and GJC performed the theoretical modelling.
\end{acknowledgments}

\bibliography{lit}

\end{document}


\title{Supplementary Material: Long-lived non-equilibrium superconductivity in a non-centrosymmetric Rashba semiconductor}

\author{V. Narayan$^{1,*}$, P. C. Verpoort$^1$, J. R. A. Dann$^1$, D. Backes$^{1}$, C. J. B. Ford$^1$, M. Lanius$^3$, A. R. Jalil$^3$, P. Sch{\"u}ffelgen$^3$, G. Mussler$^3$, G. J. Conduit$^1$, D. Gr{\"u}tzmacher$^3$}
\affiliation{
  $^1$Department of Physics, University of Cambridge, J.J. Thomson Avenue, Cambridge CB3 0HE, UK. \\
  $^2$Department of Materials Science \& Metallurgy, University of Cambridge, J.J. Thomson Avenue, Cambridge CB3 0FS, UK. \\
  $^3$Peter Gr\"unberg Institute (PGI-9), Forschungszentrum J\"ulich \& J\"ulich-Aachen Research Alliance (JARA-FIT), 52425 J\"ulich, Germany. \\
  $^\dagger$Current address: Department of Physics, Loughborough University, Loughborough LE11 3TU, UK. \\
  $^*$Correspondence to: V. Narayan $\langle$\href{mailto:vn237@cam.ac.uk}{\nolinkurl{vn237@cam.ac.uk}}$\rangle$
}

\date{\today}


\maketitle

\section{Sample growth and characterisation}

The GeTe thin film was grown by molecular beam epitaxy (MBE) on \SI{100}{\nano\meter}  Si(111) substrate, which had been wet chemically cleaned by a RCA HF-last ($\mathrm{H}_2\mathrm{SO}_4$: $\mathrm{H}_2\mathrm{O}_2$ 2:1) procedure prior to growth. The dip in $\SI{1}{\percent}$ HF passivates the surface during ex-situ transport to the MBE growth chamber. After removing the hydrogen passivation by flash annealing under UHV conditions at \SI{550}{\celsius}, the substrate temperature was set to \SI{300}{\celsius} for the growth of GeTe. Knudsen effusion cell temperatures were then set to their operating temperatures (Ge = \SI{1250}{\celsius}, Te = \SI{300}{\celsius}). The Te shutter was opened \SI{10}{\second} before the Ge shutter in order to saturate the dangling Si bonds. \SI{18}{\nano\meter} of GeTe were deposited at a growth rate = \SI{6}{\nano\meter\per\hour}.

\begin{figure}[b]
  \includegraphics[width=\textwidth]{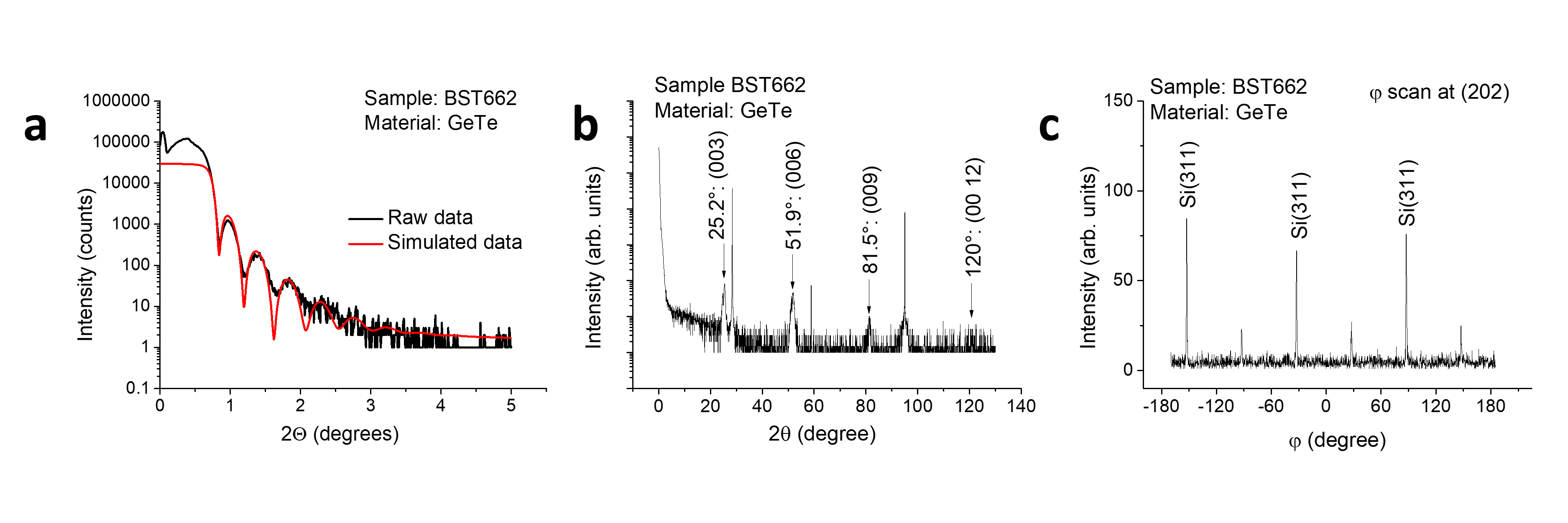}
  \caption{Thin-film characteristics. Panel (a) depicts the XRR curve of the GeTe film. Numerous oscillations are observed, and from the periodicity of these oscillations a GeTe thickness $d = \SI{18}{\nano\meter}$ is ascertained. (b) shows the $2\theta/\theta$ XRD scan. Four peaks are seen that can unambiguously be attributed to a single-crystal GeTe film with the (0,0,1) orientation in growth direction. The $\varphi$ scan shown in (c) shows 6 peaks, owing to the presence of \SI{60}{\degree} twin domains. However, three peaks show low intensity, evidencing the suppression of one domain.}
  \label{fig:fig01}
\end{figure}

X-ray diffraction (XRD) measurements were performed by means of a Bruker D8 high-resolution diffractometer utilizing the Cu $K_{\alpha1}$ wavelength ($\lambda = \SI{0.154}{\angstrom}$). The system is equipped with a 4 crystal monochromator and a G\"obel mirror in order to collimate and monochromatize the X-ray radiation. A scintillator detector was employed to collect the diffracted intensity. XRD and X-ray reflectometry (XRR) scans were performed in a symmetric $2\theta/\theta$ scans configuration with an angular resolution of \SI{0.005}{\degree} (XRR) and \SI{0.1}{\degree} (XRD). For the $\varphi$ scans at the (202) reflection of GeTe, a skew geometry was employed ($\theta = \SI{14.94}{\degree}$, $2\theta = \SI{29.89}{\degree}$, tilt angle $\xi = \SI{55.98}{\degree}$), and the scan was performed by rotating the sample by $\SI{360}{\degree}$ with a resolution of $\SI{0.1}{\degree}$.

\section{Device fabrication and electrical measurements}

The MBE films were patterned into Hall bar devices using photolithography and argon ion milling, and a subsequent stage of photolithography was used to deposit Ti/Au Ohmic contact pads.

The devices were cooled down in cryostats equipped with superconducting coils to apply a magnetic field. The data shown here are taken in a He-3 system with base temperature \SI{280}{\milli\kelvin}, and two different dilution refrigerators with base temperature \SI{50}{\milli\kelvin} and \SI{110}{\milli\kelvin}, respectively. Electrical measurements were made in a four-terminal setup using lock-in techniques at a frequency of \SI{77}{\hertz}. The excitation current used was in the range of \SIrange{10}{100}{\nano\ampere}.

When recording the data for the WAL traces shown in Fig.~1b, we were careful to maintain $|B| < \SI{0.3}{\tesla}$ at all times as this allowed us to suppress the non-equilibrium superconducting state (see \figrefsup{fig06}).

\section{Possible physical scenarios}

In this section we discuss and rule out various scenarios that could result in the reported experimental observations.

\textit{Magneto-caloric cooling:} The enhancement of superconductivity while sweeping $B$ towards \SI{0}{\tesla} might be seen as being due to adiabatic cooling of the system as spin ordering is lost. However, this is clearly ruled out by the sweep-rate dependence of $\Rxx(B)$ shown in Fig.~2e in the case of magneto-caloric cooling one would observe a more strongly superconducting behaviour for higher $|\dBbydt|$, which is inconsistent with our findings. Furthermore, under the influence of cooling one can only expect an enhancement of $\Bc$ and not the observed suppression of superconductivity at $B = \SI{0}{\tesla}$. One can imagine a more complex situation in which magnetocaloric cooling is accompanied by heating due to Eddy currents. We rule this situation out as described in \figrefsup{fig05}.

\textit{Vortex motion:} Vortex creep is a well-known phenomenon in superconductors, which may manifest as ultra-slow dynamical behaviour. While it would appear unlikely that vortices play a role in the observed phenomena since the slow dynamics are seen well above $\Tc$, nevertheless, to investigate the possibility of some degree of phase rigidity above \SI{0.1}{\kelvin} we conducted magnetisation ($M$) measurements. \figrefsup{fig06} shows that
within experimental error there is no $T$-variation in the $M(B)$ traces between \SI{0.4}{\kelvin} and \SI{1.0}{\kelvin}. This strongly suggests the absence of any phase rigidity and/or superconducting vortices at \SI{0.4}{\kelvin}.

\textit{Nuclear spin dynamics:} The dynamics of nuclear spins is another source of slow relaxation in low-$T$ condensed matter systems. However, the Overhauser field due to nuclear spins only depends on $B$ and not $\dBbydt$~\cite{Kumada_etal_PRL2007, Chekhovic_etal_NatMat2013}, which is inconsistent with our observations. Furthermore, the Overhauser field cannot account for the asymmetry between $\Rxx(\pm B)$ for a given sweep, or the mirror symmetry between the forward and backward $B$ sweeps.

\textit{Trapped flux in the superconducting magnet:} Trapped flux in the superconducting coil would also manifest as an offset in $B$ and so can be ruled out for the same reasons as nuclear spins. Furthermore, the observed behaviour would indicate a diamagnetic correction to the external field (since the minimum in $\Rxx$ appears before crossing $B = \SI{0}{\tesla}$), which is clearly inconsistent with the expected corrections due to the external superconducting magnet.

\begin{figure}[h]
  \includegraphics[width=\textwidth]{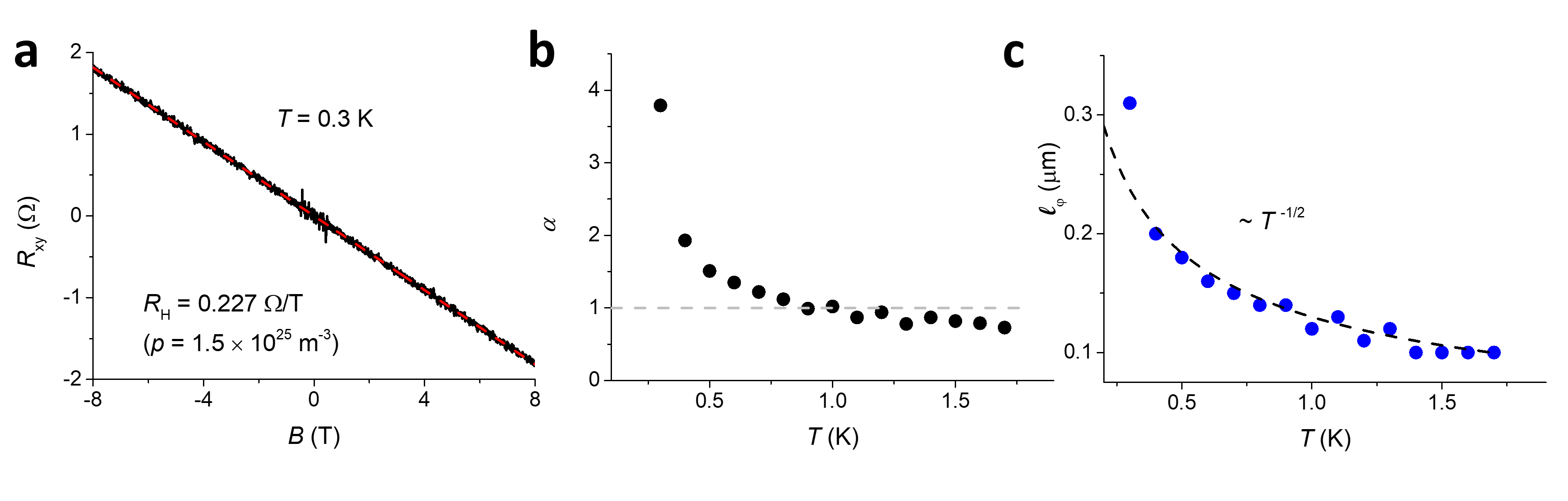}
  \caption{Mobility characteristics and evidence of 2D nature. Panel (a) shows the Hall characteristics of the GeTe which correspond to a 3D carrier density of $\SI{1.5e+25}{\per\cubic\meter}$. However, panels (b) and (c) which display the HLN fit parameters would suggest that the transport is mediated by 2D channels. The parameter $\alpha$ in the HLN formula~\cite{Hikami80} assumes a value $0.5$ for each 2D WAL mode. At $T > \SI{0.5}{\kelvin}$ the data is consistent with two surface modes (the enhanced $\alpha$ at lower $T$ is reflective, presumably, of the onset of superconducting correlations~\cite{Narayan16}). Panel (c) shows that the phase coherence length $l_\varphi$ decays as $T^{-\nicefrac{1}{2}}$ which is consistent with the dephasing behaviour due to inter-carrier interactions in 2D.}
  \label{fig:fig02}
\end{figure}

We also mention here that the Hall signal (\figrefsup{fig02}a) of our samples rules out the existence of any stray magnetic fields (due to magnet power supply, from electrical contacts~\cite{Svanidze15} etc.) since neither the asymmetry between $B$ and $-B$, nor the rate dependence is reflected in $\Rxy$.

\textit{Inhomogeneity effects:} $\Tc$ in GeTe depends on the concentration of carriers~\cite{Hein64}, which in our high-quality molecular-beam-epitaxy (MBE)-grown sample are most likely mobile Ge vacancies~\cite{Edwards05,Edwards06}. Therefore, it is plausible that the vacancy concentration and hence $\Tc$ vary spatially. However, while such a scenario might explain the relatively gradual superconducting transition in Fig.~1c, it is inconsistent with how $\Tc$ depends on both the sign and magnitude of $\dBbydt$.

\begin{figure}[p]
  \includegraphics[width=10cm]{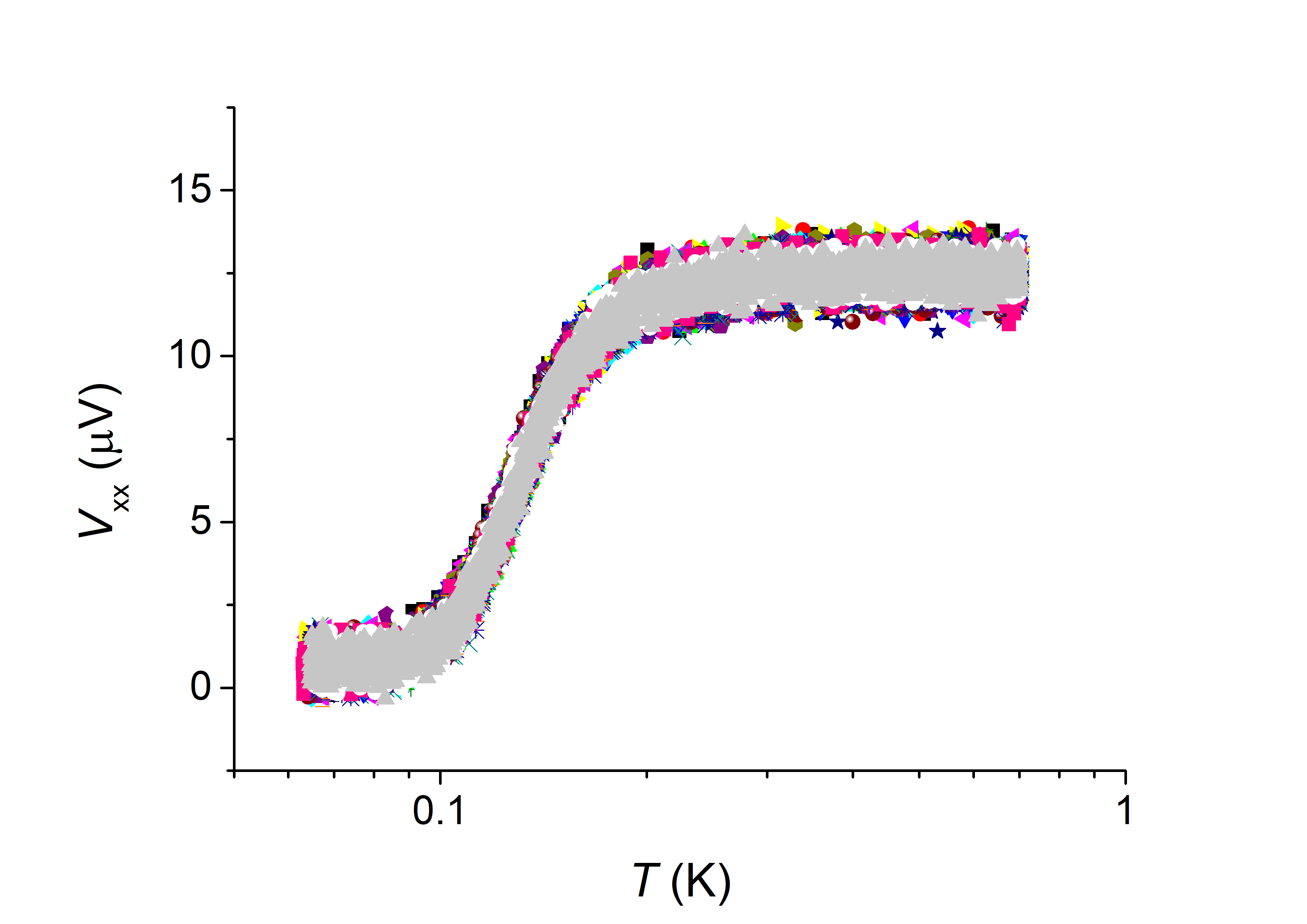}
  \caption{The figure shows the raw data used to generate Fig.~1c. The data suggests a small positive voltage offset of $\lesssim$~\SI{1}{\micro\volt} which is possibly responsible for the apparent finite $\Rxx$ in the superconducting state.}
  \label{fig:fig12}
\end{figure}

\begin{figure}[p]
  \includegraphics[width=16cm]{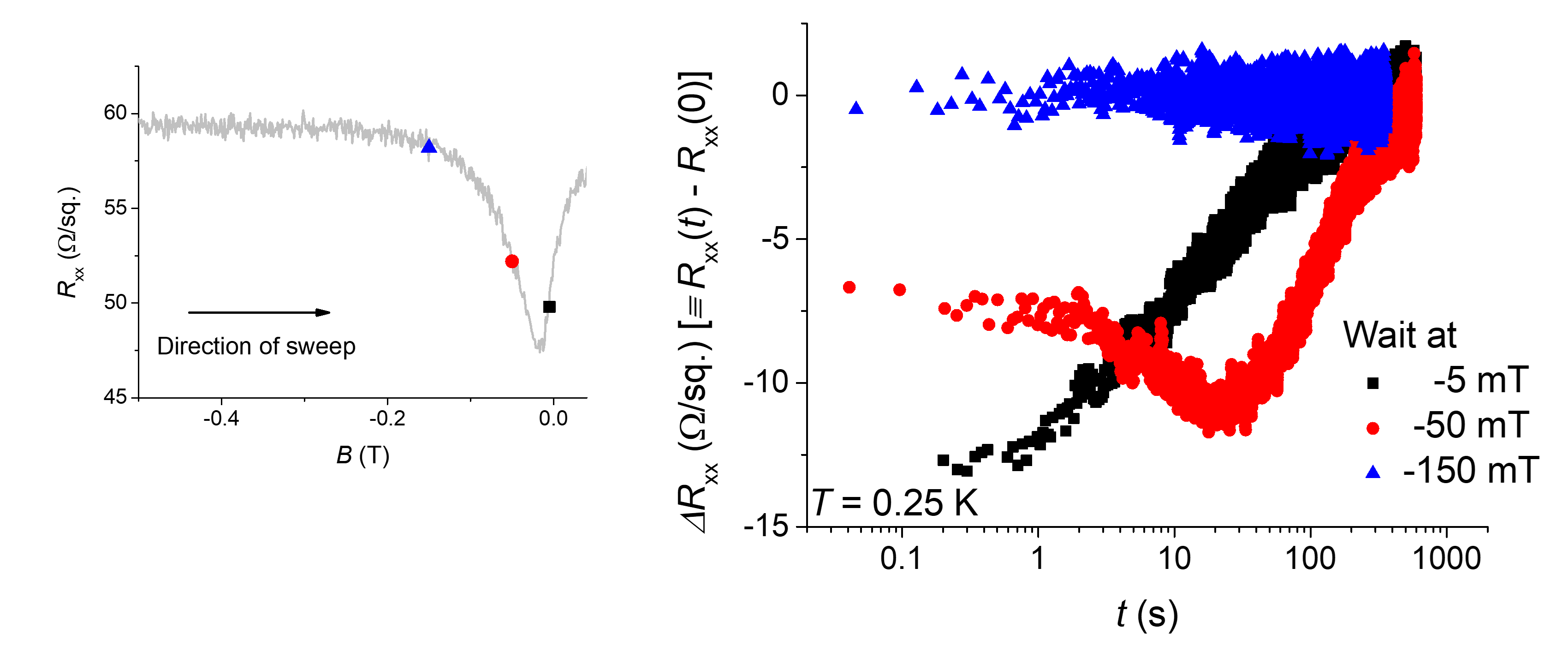}
  \caption{Slow relaxation of the non-equilibrium superconducting state. The right panel shows the time evolution of $\Rxx$ when the field-sweep is stopped at three different values of $B$ indicated on the left panel. At \SI{-150}{\milli\tesla} the state remains normal, whereas at \SI{-5}{\milli\tesla} the system relaxes to $\RN$ roughly in an exponential manner. If, however, the field sweep is stopped at $\SI{-50}{\milli\tesla}$ $\Rxx$ goes through a minimum before relaxing roughly exponentially. Similar to Fig.~2d the system relaxes fully on a timescale of several minutes.}
  \label{fig:fig03}
\end{figure}

\begin{figure}[p]
  \includegraphics[width=\textwidth]{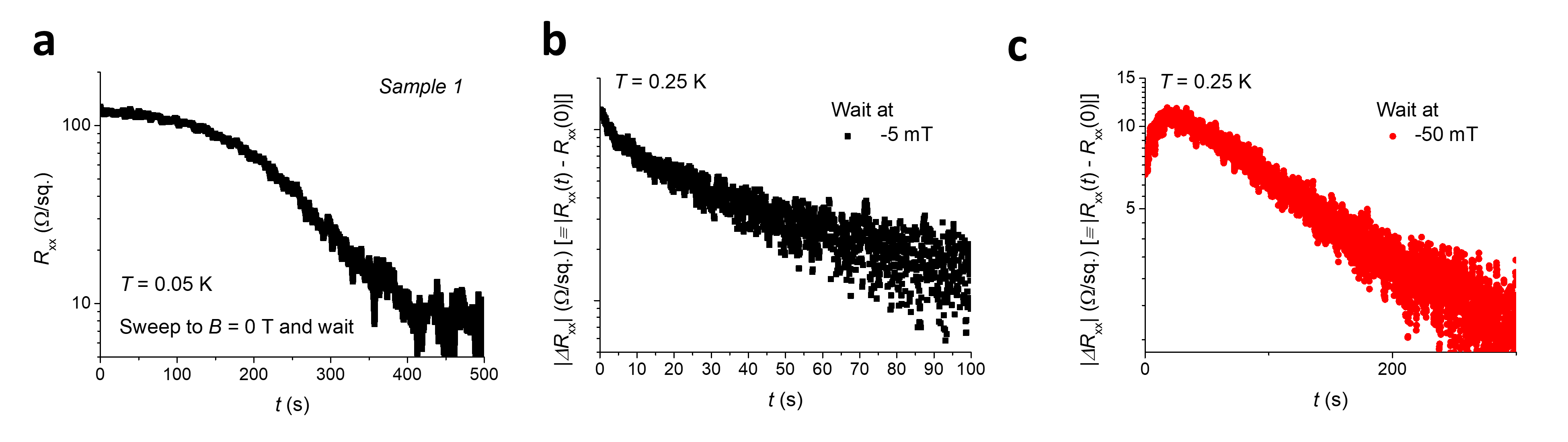}
  \caption{The panels show data from Fig.~2d and \figrefsup{fig03} on a log-linear scale. In each instance the long-time behaviour is consistent with an exponential decay.}
  \label{fig:fig11}
\end{figure}

\begin{figure}[p]
  \includegraphics[width=\textwidth]{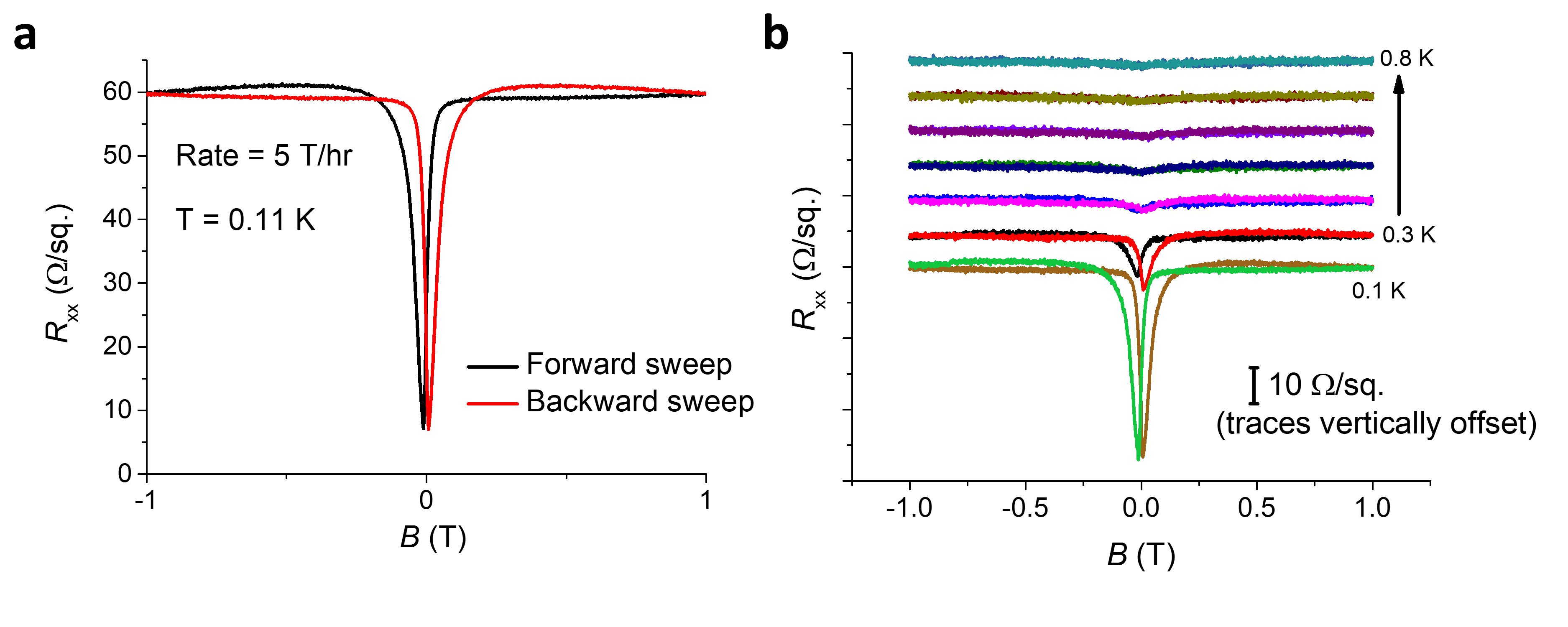}
  \caption{Non-equilibrium superconducting state evidenced in a second sample cooled in a separate cryostat with base temperature = \SI{0.11}{\kelvin} where, under equilibrium conditions, the sample did not go fully superconducting ($R_{\mathrm{s}}/\RN \approx 0.75$). Several field sweeps have been averaged to produce the data in panel (a). In panel (b) we show (non-averaged) data indicating the presence of the non-equilibrium superconducting state up to \SI{0.4}{\kelvin}.}
  \label{fig:fig04}
\end{figure}

\begin{figure}[p]
  \includegraphics[width=\textwidth]{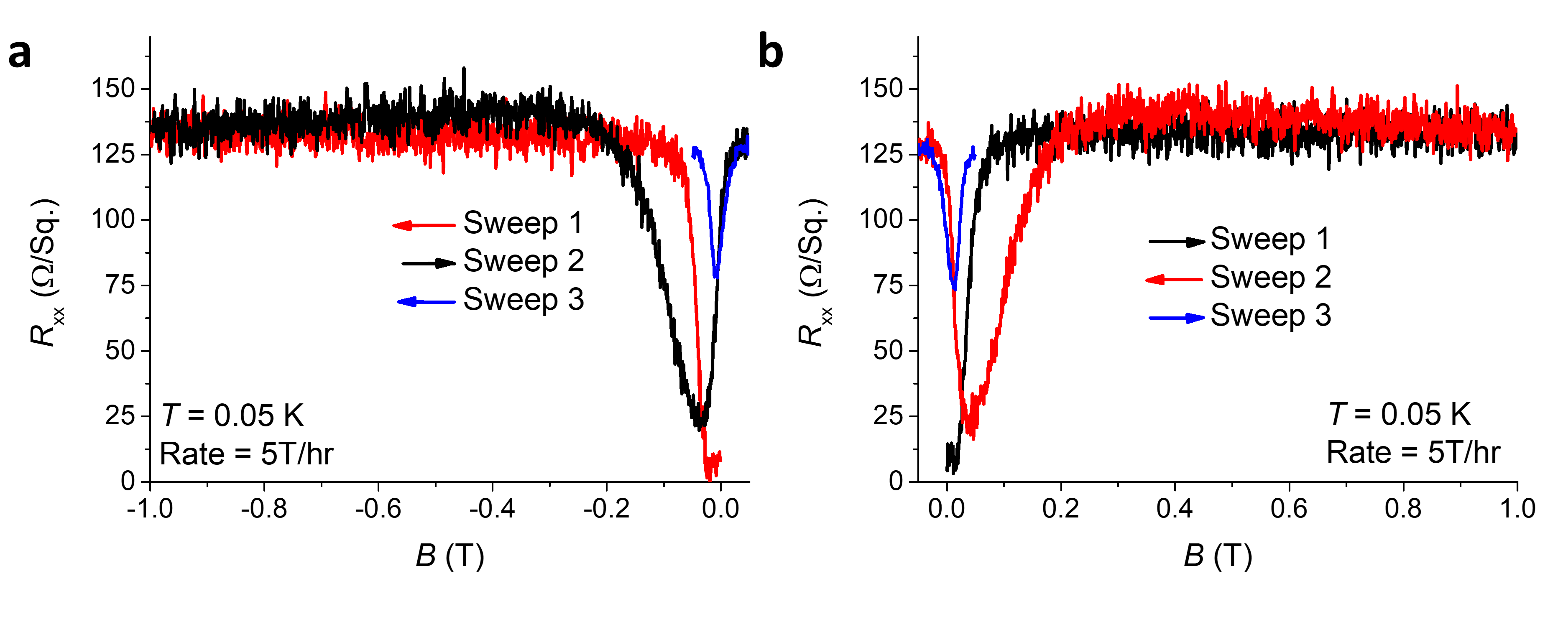}
  \caption{Absence of Eddy current heating. (a) The above experiment investigates whether the superconducting state at $B = \SI{0}{\tesla}$ in sweep 2 is not recovered because of Eddy current heating. The scenario under question is as follows: starting from the superconducting state, $B$ is swept to \SI{-1}{\tesla}. In sweep 2, below $|B| = \SI{0.25}{\tesla}$ the sample is cooled by the magnetocaloric effect, but then heat generated due to Eddy currents overwhelms this cooling and causes the sample temperature to rise above $\Tc$ ultimately causing the superconducting state at \SI{0}{\tesla} not to manifest. In the experiment shown above, sweep 2 is stopped at \SI{0.05}{\tesla} and the sweep direction is reversed immediately. If the sample temperature was indeed above $\Tc$, then a lowering of $\Rxx$ is not expected for several minutes (Fig.~2d). However, we find a clear drop in $\Rxx$ which indicates that the sample temperature is not significantly raised. Not only does this counter the possibility of Eddy current-induced heating, it is further evidence against magnetocaloric cooling, since at $B = \SI{0.05}{\tesla}$, one doesn't expect significant ordering of spins. (b) The situation is precisely mirrored if the sweeps are performed in the opposite direction.}
  \label{fig:fig05}
\end{figure}

\begin{figure}[p]
  \includegraphics[width=10cm]{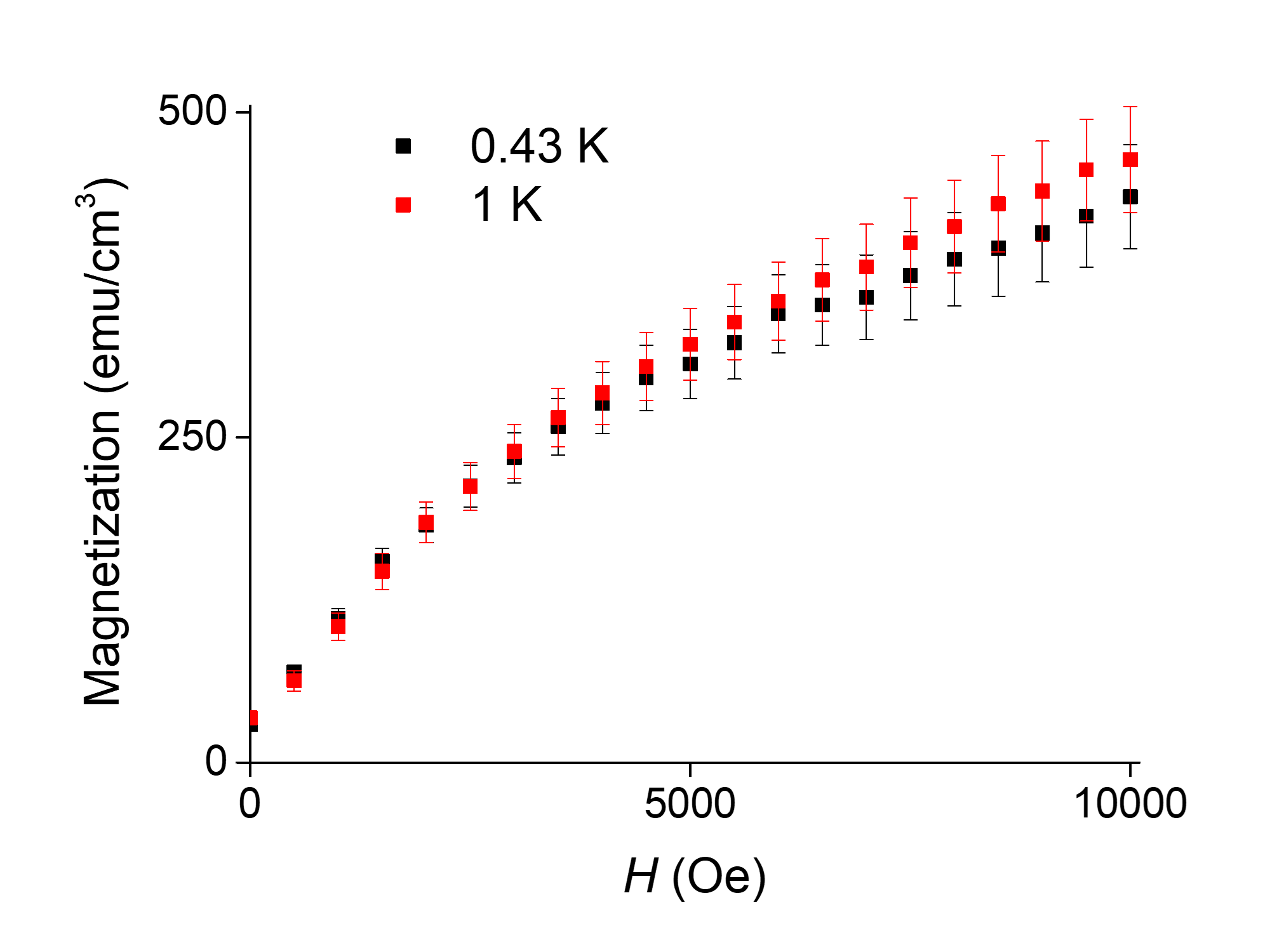}
  \caption{Low-$T$ equilibrium magnetisation $M(H)$ characteristics. The dependence of $M(H)$ is manifestly paramagnetic showing no sign of diamagnetism associated with superconductivity. While we note that in the thin-film samples it might be expected that the Si substrate contributes substantially to the measured signal, we note that within experimental error, there is no difference in the traces measured at \SI{0.43}{\kelvin} and \SI{1}{\kelvin}. This indicates that there is no phase rigidity at \SI{0.43}{\kelvin} under equilibrium conditions. The data is collected by incrementing the field in small amounts and allowing any transients to pass before recording. The volume of the sample is $\SI{1}{\centi\meter} \times \SI{0.3}{\centi\meter} \times \SI{18}{\nano\meter}$.}
  \label{fig:fig06}
\end{figure}

\begin{figure}[p]
  \includegraphics[width=\textwidth]{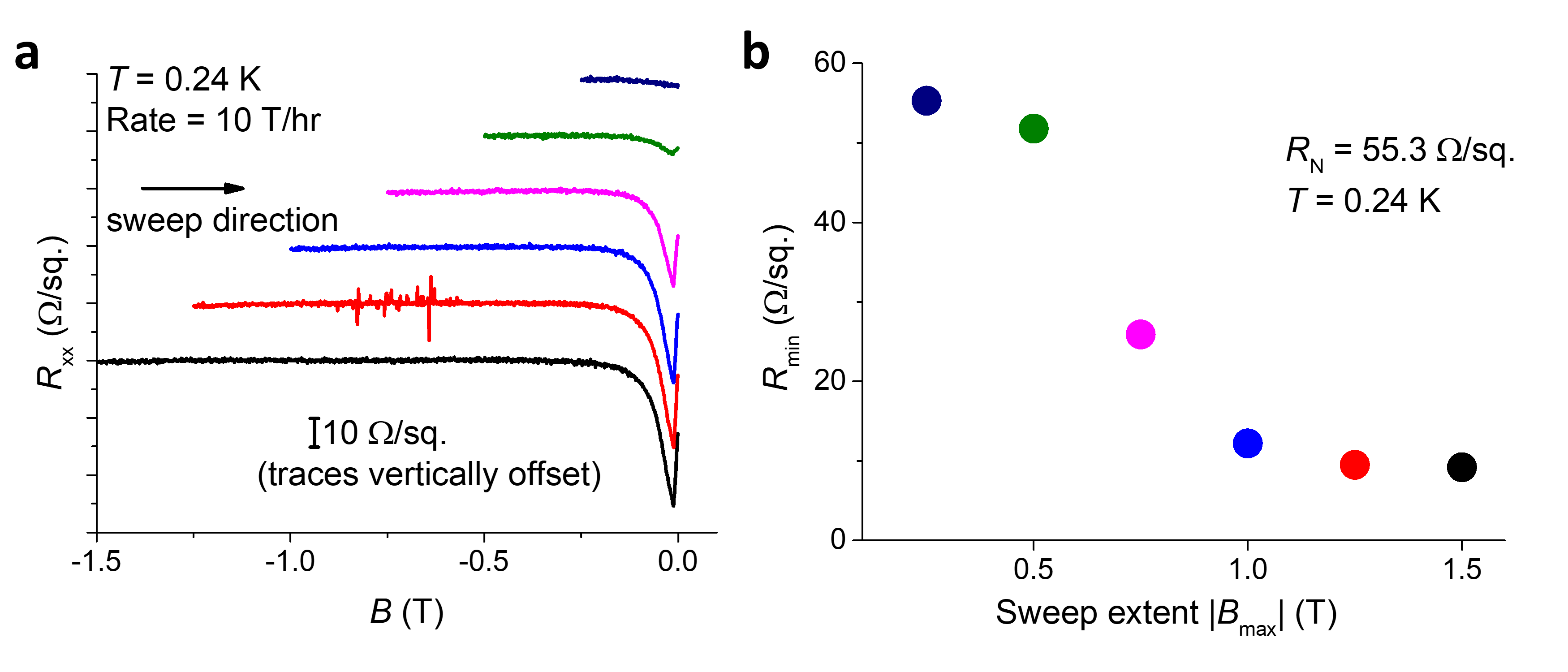}
  \caption{The figure shows the dependence of the non-equilibrium superconducting state on the `sweep extent` $\Bmax$, i.e., the maximum value of $|B|$ swept to. Each experiment is started under equilibrium conditions at $B = \SI{0}{\tesla}$ (in contrast to \figrefsup{fig05}) and only the sweep from $\Bmax$ to \SI{0}{\tesla} is shown. The right panel shows the minimum resistance Rmin as a function of $|\Bmax|$. The occurrence of the non-equilibrium superconducting state depends strongly on whether a large enough Zeeman field is applied. As shown in the figure, for absolute field values less than \SI{0.5}{\tesla}, the nonequilibrium behaviour is not induced. This is expected since for $B < \SI{0.5}{\tesla}$ the thermal energy ($\kB T$) is comparable to the Zeeman energy ($g \muB B$), and consequently, the population imbalance between the Rashba bands is offset by the thermal smearing of the Fermi function.}
  \label{fig:fig07}
\end{figure}

\bibliography{lit}